\documentclass[11pt]{article}
\usepackage{xspace}
\usepackage{graphicx}
\usepackage{amsmath}
\usepackage{amssymb}
\usepackage{color}
\usepackage{caption}
\usepackage{subcaption}
\usepackage{url}
\usepackage{lineno}

\definecolor{Red}{rgb}{1,0,0}
\definecolor{Green}{rgb}{0,1,0}
\definecolor{Blue}{rgb}{0,0,1}
\definecolor{Black}{rgb}{0,0,0}

\def\beq{\begin{equation}}
\def\eeq#1{\label{#1}\end{equation}}
\def\eeqn{\end{equation}}

\def\beqa{\begin{eqnarray}}
\def\eeqa#1{\label{#1}\end{eqnarray}}
\def\eeqan{\end{eqnarray}}

\let\bar=\overbar

\def\etal{{\it et al.}}

\def\Dslash{\not{\hbox{\kern-4pt $D$}}}
\def\dslash{\not{\hbox{\kern-2pt $\del$}}}

\def\msb{{\bar{\ssstyle M \kern -1pt S}}}

\def\Title#1{\begin{center} {\Large {\bf #1} } \end{center}}

\begin{document}

\Title{SUSY searches with the ATLAS detector }

\bigskip\bigskip


\begin{raggedright}  

Lucian-Stefan Ancu\index{Ancu, L.S.}, {\it DPNC, Universite de Geneve}\\

\begin{center}\emph{On the behalf of the ATLAS Collaboration.}\end{center}
\bigskip
\end{raggedright}

{\small
\begin{flushleft}
\emph{To appear in the proceedings of the Interplay between Particle and Astroparticle Physics workshop, 18 -- 22 August, 2014, held at Queen Mary University of London, UK.}
\end{flushleft}
}

\section{Introduction}
\label{sec:introduction}

Supersymmetry(SUSY) -- for a general introduction see Ref. \cite{Martin:1997ns} -- is an appealing extension of the Standard Model (SM).  The newly discovered Higgs  \cite{Aad:2012tfa,Chatrchyan:2012ufa} at a mass of 125 GeV is fully compatible with many of the  SUSY models,  where one  expects  a light Higgs boson~(h). 

SUSY can be R-parity\footnote{R parity is defined as $\textrm{R-parity}=(-1)^{(3(B-L)+2s)}$ where B is the baryon number,L the lepton number and s the spin.  For SM particles the parity is equal to +1 while for SUSY particles it is -1. } conserving or violating. In the conserving scenarios the Lightest Supersymmetric Particle (LSP) is the lightest neutralino ($\tilde\chi_1^0$) that is a good  candidate for the dark matter particle content.   The ATLAS  experiment  has performed a wide variety of searches for SUSY in both the 7 TeV and 8 TeV datasets searching for both R-conserving and R-violating  models.  In the following only a few examples of the ATLAS searches will be highlighted the full list being available at Ref.\cite{ATLAS:susyresults}.

\begin{figure}[!hb]
\begin{center}
\includegraphics[width=0.5\columnwidth]{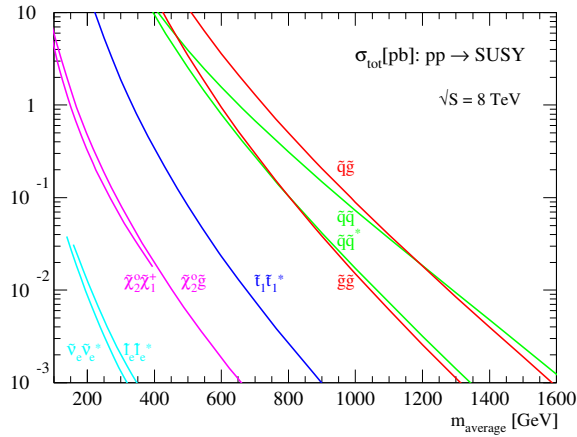}
\caption{Cross section for the production of sparticles at LHC at 8 TeV center of mass energy as a function of the average mass of the pair produced. \cite{Prospino:plot,Beenakker:1996ch,Beenakker:1997ut,Beenakker:1999xh} }
\label{fig:xsections}
\end{center}
\end{figure}


The ATLAS detector  is a general purpose detector and a full description and  its expected performance can be consulted in Ref. \cite{ATLAS:1999uwa}. In these proceedings I will report on analyses performed on data collected by the ATLAS detector in  4.7 $\rm{fb^{-1}}$ at a center-of-mass energy $\sqrt s  = \textrm{7 TeV}$  and 20.3 $\rm{fb^{-1}}$ at  $\sqrt s =\textrm{8 TeV}$. 
In the SUSY searches the full capabilities of the ATLAS detector are used, as the targeted  signatures include all reconstructed physics objects: muons, electrons, taus, photons, jets, b-tagged jets, transverse missing energy. A good understanding of the SM processes is compulsory because the signal events are, many times, expected to be located in tails of distributions or as  deviations over small SM event yields. As no excess is seen over the SM expected yields, limits on the SUSY particles masses are set. 

\section{ATLAS SUSY searches strategy}
\label{sec:strategies}

With the many parameters (105) introduced by the Minimal Supersymetric Model (MSSM), additional constraints are needed when   searching for SUSY. In the last years the strategy has moved towards what is named Natural SUSY. In Natural SUSY the third generation and elecroweakinos are light, while the other supersymmetric particles can be heavy. The cross-section for the direct  production of supersymmetric particles at the LHC can be seen in Fig.\ \ref{fig:xsections}. 

\begin{figure}[!t]
        \centering
        \begin{subfigure}[b]{0.55\textwidth}
                \includegraphics[width=\textwidth]{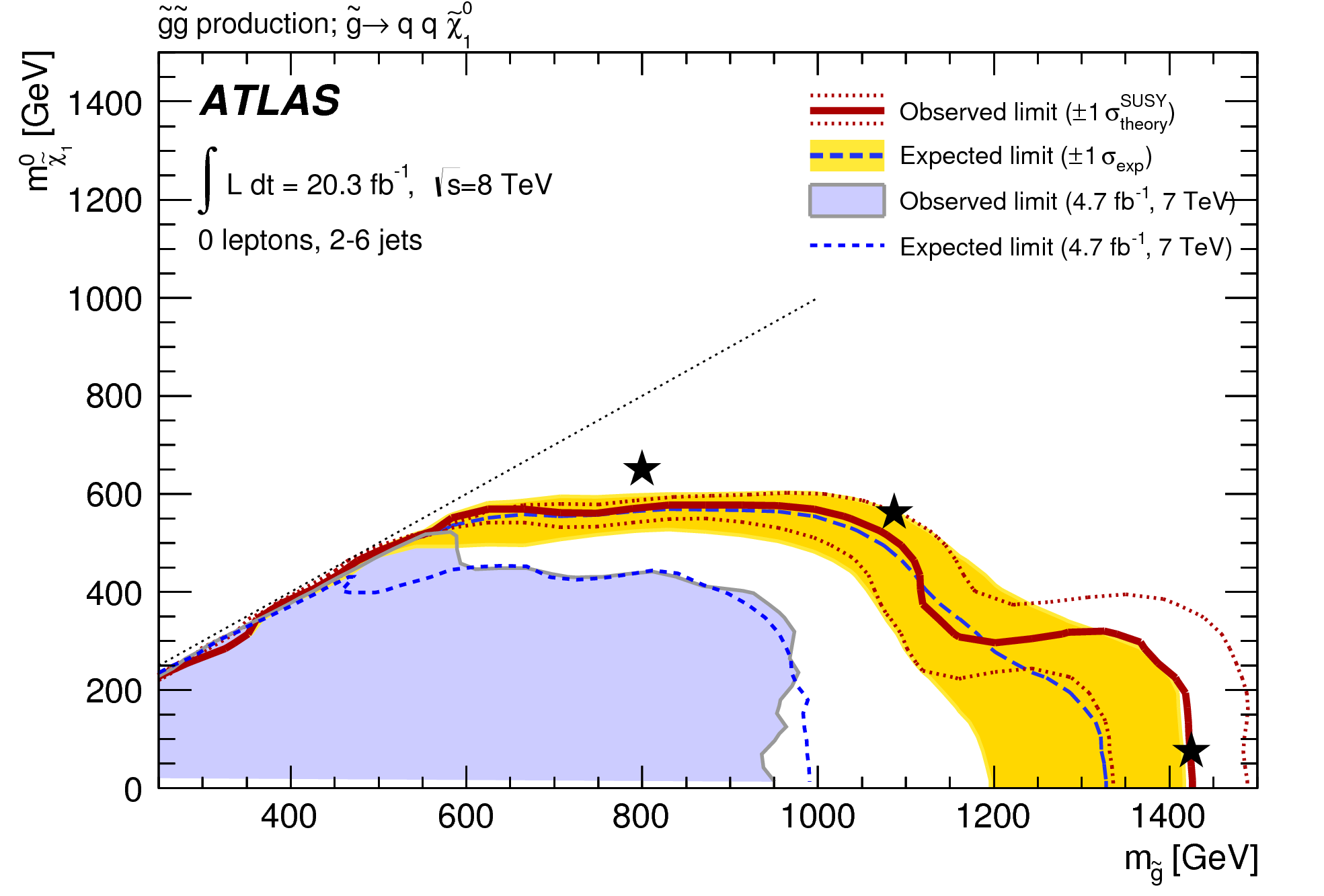}
                \caption{}
                \label{fig:gluino}
        \end{subfigure}
                \begin{subfigure}[b]{0.4\textwidth}
                \includegraphics[width=\textwidth]{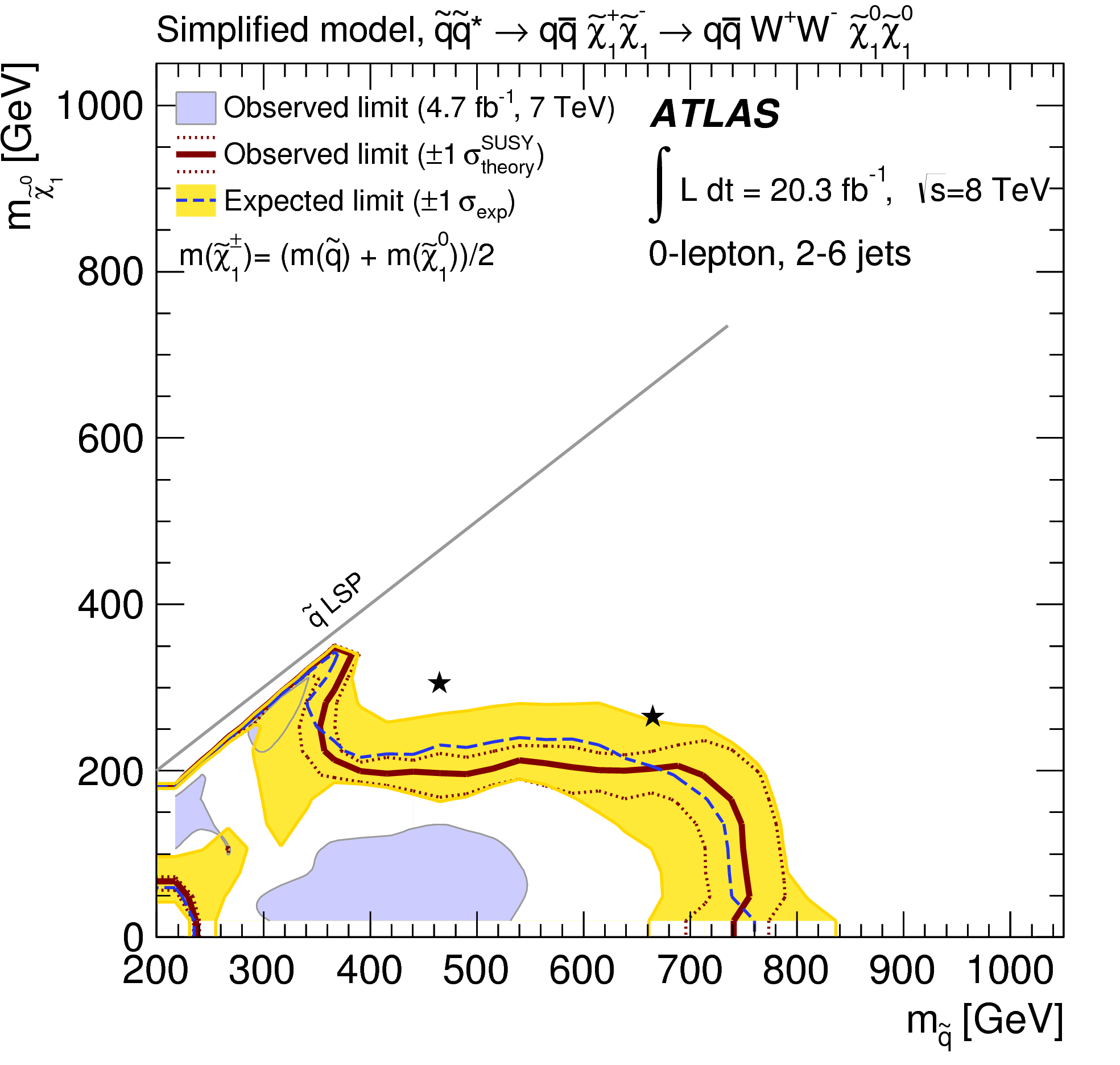}
                \caption{}
                \label{fig:squark_boosted}
        \end{subfigure}%
                \caption{(a)Exclusion limits for the pair production of gluinos. (b) Exclusion limits for squarks decaying via charginos.  The dashed and solid lines show the expected and observed limits, respectively, including all uncertainties except the theoretical signal cross section uncertainties.   }\label{fig:squarks_gluino}
\end{figure}

Based on the type of  decay  the produced supersymmetric particles can decay promptly or be long lived. Most of the ATLAS searches concentrate on searches for promptly decaying supersymmetric particles but in Section \ref{sec:longlived} an example of a search for long lived particles is presented.  As the realisation of SUSY in nature can be R-parity conserving or violating this fact is taken into account both in the production and decay of the searched particles. 

When considering full R-parity conserving models the LSP ( $\tilde\chi_1^0$) will be produced in pairs in cascade decays of directly produced sparticles. The signature of the LSP is the presence of additional missing energy. Due to the cascade nature of the decays of SUSY particles a large multiplicity  of jets is expected, especially in decays involving coloured particles. SUSY processes are expected to produce deviations from the SM expected distributions especially at high missing transverse energy, large jet multiplicity and in some variables that reconstruct properties of the hypothesised SUSY particle.

In the following searches for squars, gluinos, third generation squarks and electroweakinos in R-parity conserving scenarios are presented followed by an example of searches in R-parity violating scenarios. 

\subsection{Search for squarks and gluinos}
\label{sec:squarks}

Squarks ($\tilde q$) and gluinos ($\tilde g$) are expected to be produced in pairs ($\tilde q \tilde q $, $\tilde q \tilde g $, $\tilde g \tilde g $) in many of the R-parity SUSY models. Some of the most simple  decays are to neutralinos ($\tilde\chi^0_1$)  $\tilde q \rightarrow q \tilde\chi^0_1$, $\tilde g \rightarrow g  \tilde\chi^0_1$  or to charginos ($\tilde\chi^\pm$) $\tilde q \rightarrow q \tilde\chi^\pm$, $\tilde g \rightarrow q\bar q  \tilde\chi^\pm$ with the charginos decaying via a $W^\pm$ to neutralinos.  Depending on the SUSY spectrum below the gluino/squark mass the chain of decays can be long, including 3-4 decay steps. Hence these decays are characterised by large jet multiplicities  in the final states.  For the simplest scenario of the gluino decay to a pair of quarks and a neutralino, gluino masses below 1.4 TeV are excluded at 95\% CL  as seen Fig.~\ref{fig:gluino} \cite{Aad:2014wea}. In Fig.\ \ref{fig:squark_boosted} the limits on the squark mass  for the scenario when the  W is boosted and reconstructed in a single jet are presented; also in this scenario squarks with a mass below 700 GeV are excluded\cite{Aad:2014wea}.

\subsection{Searches for third generation of squarks}
\label{sec:stops}

\begin{figure}[!hb]
\begin{center}
\includegraphics[width=0.6\columnwidth]{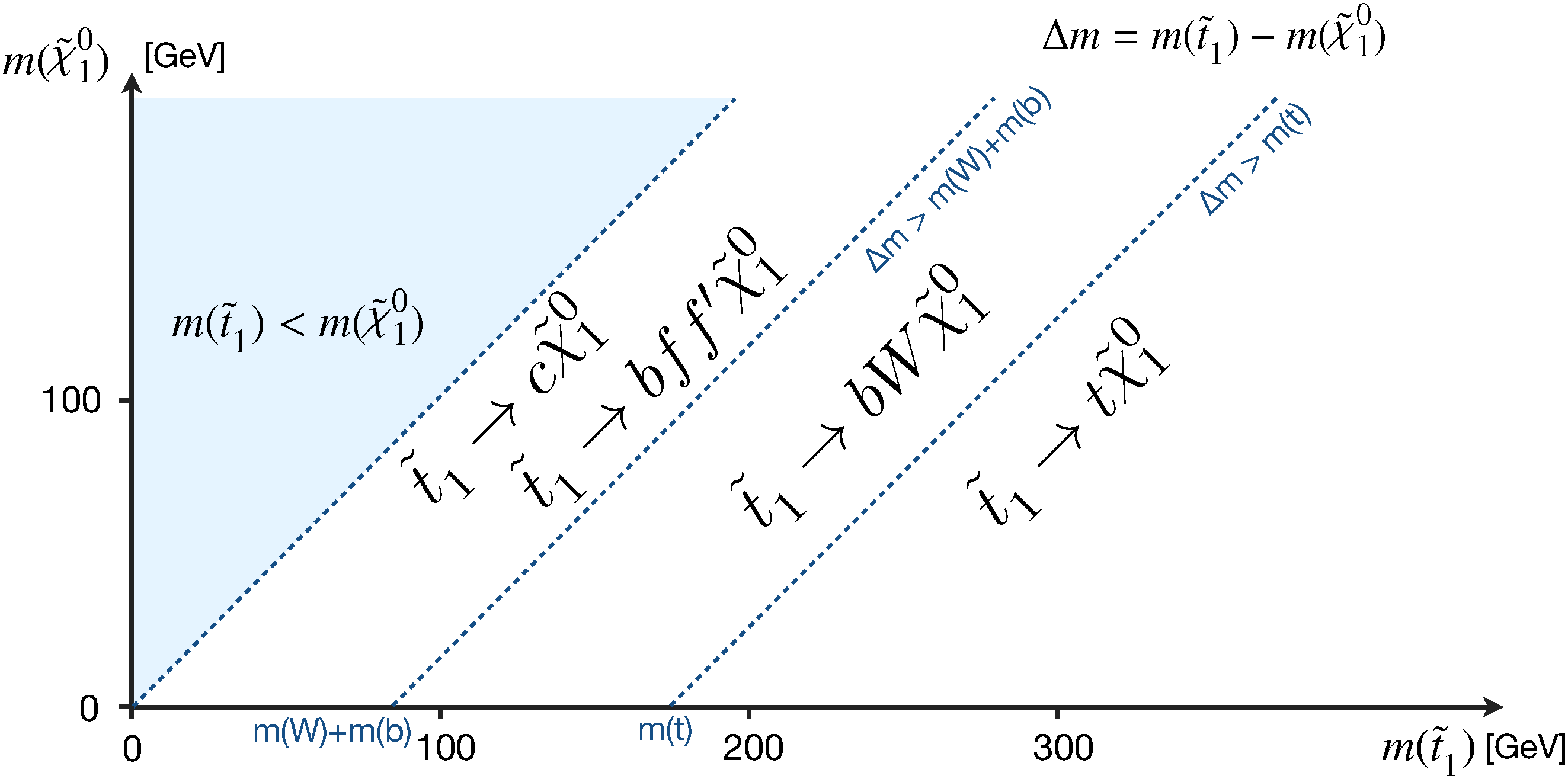}
\caption{Diagram presenting the decay modes of the stop in the plane spanned by the mass of the stop and the mass of the LSP.  The mass relations are listed in the plots\ \cite{Aad:2014kra}.}
\label{fig:stop_decays}
\end{center}
\end{figure}

\begin{figure}
\begin{center}
\includegraphics[width=0.8\columnwidth]{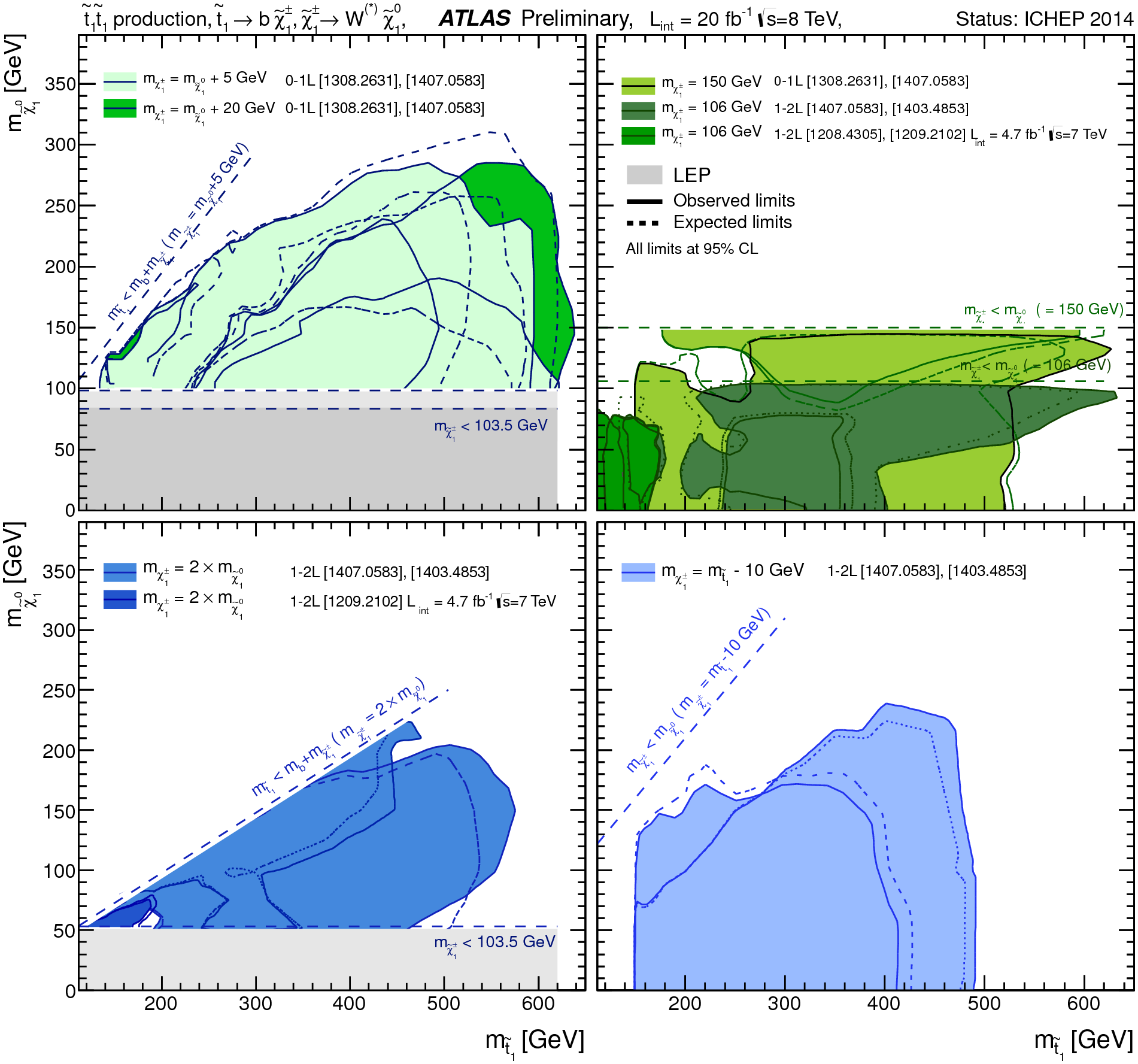}
\caption{ Expected and observed 95\% CL excluded region in the stop neutralino mass plane with different assumptions, on the stop, chargino and neutralino mass differences. The dashed and solid lines show the expected and observed limits, respectively, including all uncertainties except the theoretical signal cross section uncertainties. \cite{ATLAS:summaryplots}}
\label{fig:stop_chargino}
\end{center}
\end{figure}

\begin{figure}
        \centering
        \begin{subfigure}[b]{0.55\textwidth}
                \includegraphics[width=\textwidth]{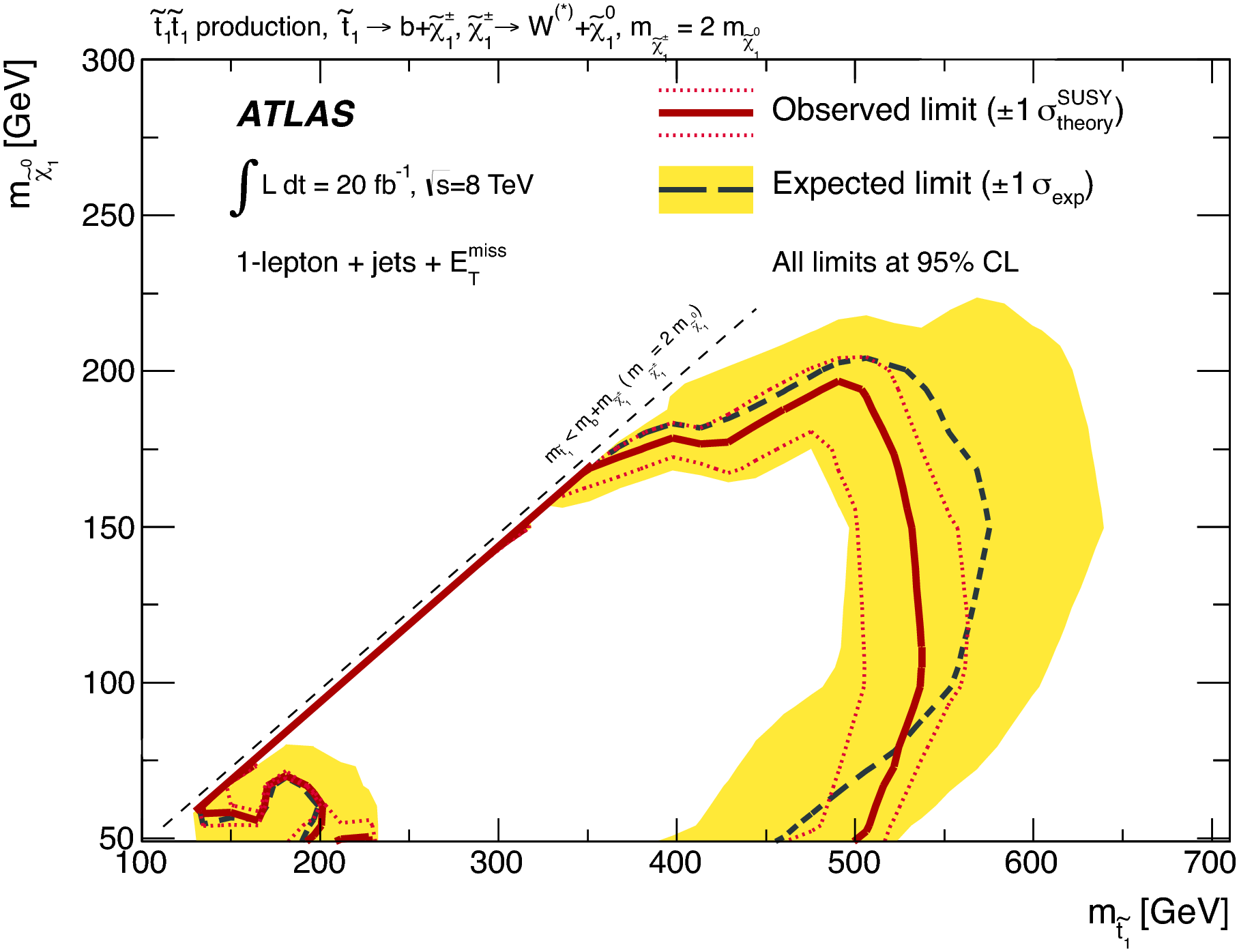}
                \caption{  }
                \label{fig:stop_bw}
        \end{subfigure}~
                \begin{subfigure}[b]{0.45\textwidth}
                \includegraphics[width=\textwidth]{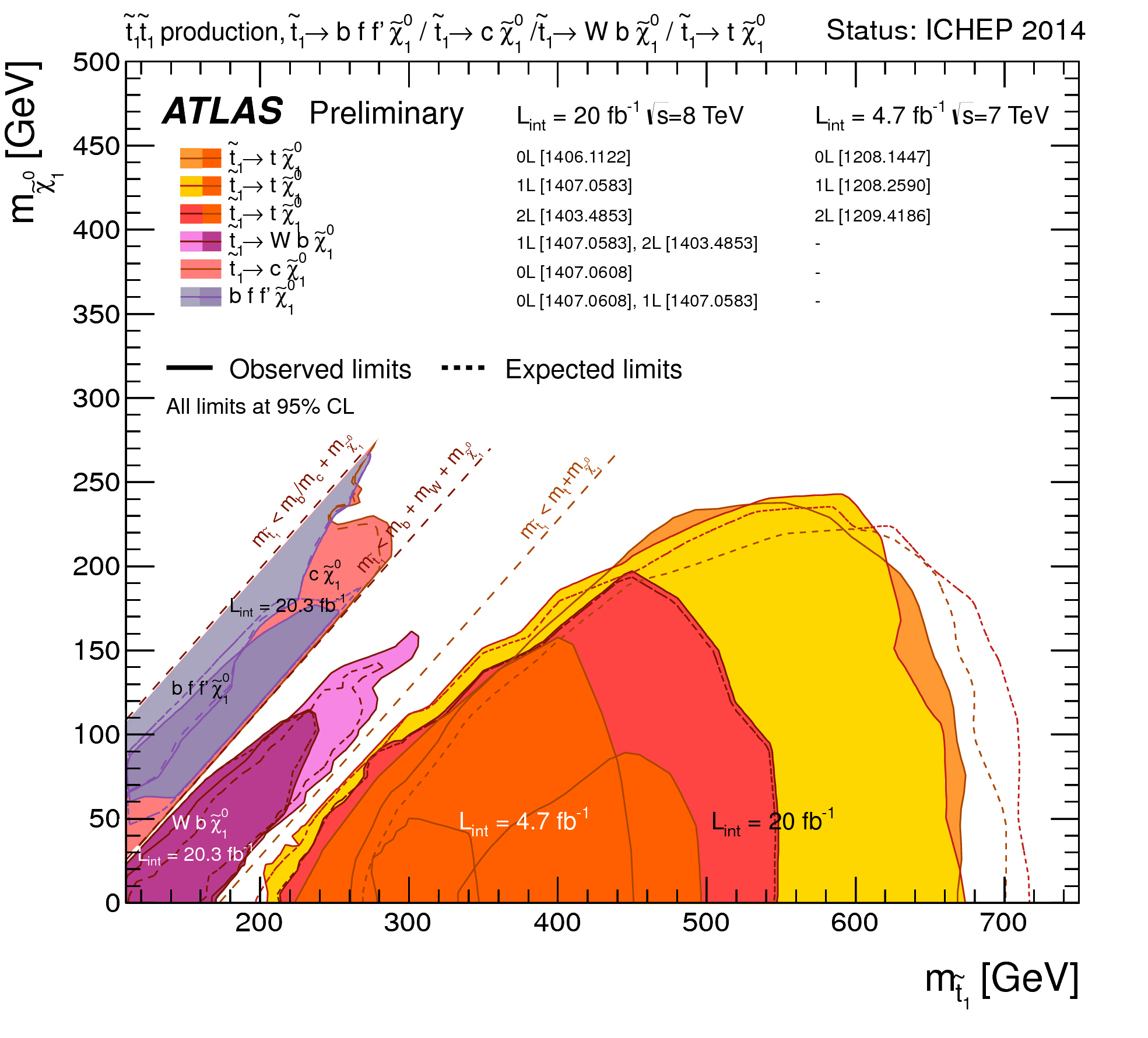}
                \caption{}
                \label{fig:stop_sum}
        \end{subfigure}%
       
        \caption{(a) Expected and observed 95\% CL excluded region in the stop neutralino plane assuming a $BR(\tilde t_1 \rightarrow b \tilde\chi_1^\pm)=100\%$ and $BR(\tilde\chi_1^\pm \rightarrow W\tilde\chi_1^0)=100\%$. The dashed and solid lines show the expected and observed limits, respectively, including all uncertainties except the theoretical signal cross section uncertainties.~\cite{Aad:2014kra}\\  (b)  Summary of the stop ATLAS searches. Exclusion limits at 95\% CL are shown in the stop neutralino mass plane. The included decay modes, all considered having $BR=100\%$ are: $\tilde t_1 \rightarrow  t \tilde\chi_{1}^{0}$, $\tilde t_1 \rightarrow  bW \tilde\chi_{1}^{0}$, $\tilde t_1 \rightarrow  bff' \tilde\chi_{1}^{0}$ and $\tilde t_1 \rightarrow  c \tilde\chi_{1}^{0}$. The dashed and solid lines show the expected and observed limits, respectively, including all uncertainties except the theoretical signal cross section uncertainties.~\cite{ATLAS:summaryplots}}
\end{figure}

In Natural SUSY the superpartner of the  top quark, $\tilde t_1 $,  is  expected to have a mass below 1 TeV. Squarks of the 3rd family (stops and sbottoms) are expected to be light, because for low masses the top loop diagrams contribution to the Higgs mass can be cancelled without introducing an excessive amount of fine-tuning. Depending on the mass of the stop the following decays could be dominant:  $\tilde t_1 \rightarrow  t \tilde\chi_{1}^{0}$, $\tilde t_1 \rightarrow  bW \tilde\chi_{1}^{0}$, $\tilde t_1 \rightarrow  bff' \tilde\chi_{1}^{0}$ or $\tilde t_1 \rightarrow  c \tilde\chi_{1}^{0}$ as seen in Fig.~\ref{fig:stop_decays}. The searches are designed such that they cover all the possible decays of the stop into a neutralino LSP and make use of advanced techniques for reconstructing the decay products. For the latter case searches include topologies in which the W is boosted and reconstructed as a single jet \cite{Aad:2014kra}. Limits on the production of stop when considering the decay $\tilde t_1 \rightarrow  bW \tilde\chi_{1}^{0}$ can be seen in Fig.~\ref{fig:stop_bw} and  in  all considered  scenarios can be seen in Fig. \ref{fig:stop_sum}. For a massless $\tilde\chi_{1}^{0}$ the stop can be excluded up to 650 GeV (except some regions where the mass difference between the stop and the neutralino is near  the top mass). Details on the limits when the decay of the stop goes via a chargino, in different mass scenarios, can be seen in Fig. \ref{fig:stop_chargino} and also in these scenario the stop is excluded for masses up to ~400 GeV. 

\subsection{Electroweak sector searches}
\label{sec:ew}

Direct production of charginos ( $\tilde\chi^\pm$) and neutralinos ($\tilde\chi^{0}_i$), considering the $\tilde\chi^{0}_1$ as LSP, is expected to have a clean signature due to the presence of SM bosons and  leptons in the decay chains. Depending on the mass splitting a heavier neutralino can decay to the LSP via a Z or the SM Higgs, and the charginos via a W. The final states in direct pair production of charginos and neutralinos are characterised by a large lepton multiplicity (~$\geq 2$).  Depending on the decay, a dilepton invariant mass compatible with the Z mass can be vetoed or required. 

\begin{figure}
        \centering
        \begin{subfigure}[b]{0.5\textwidth}
                \includegraphics[width=\textwidth]{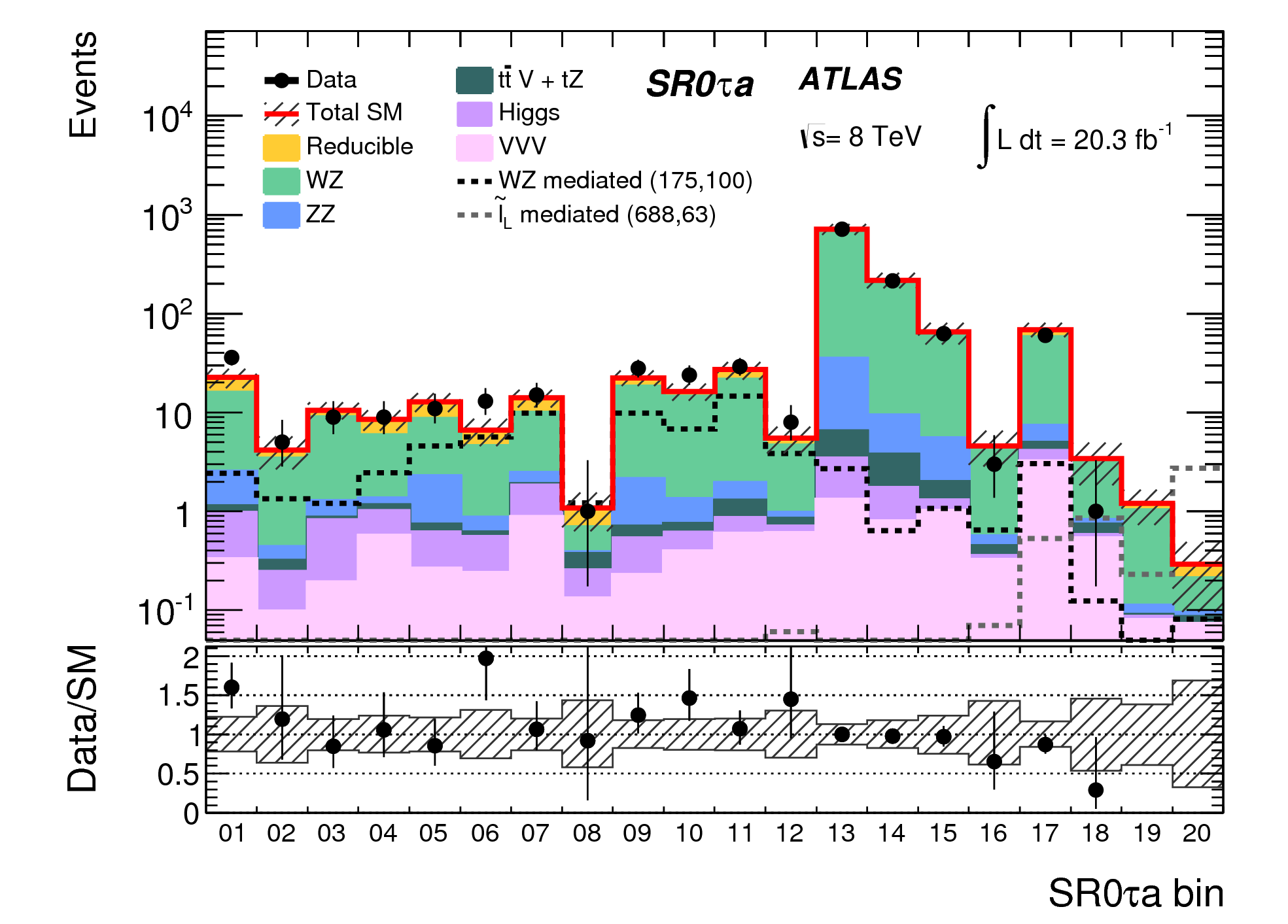}
                \caption{}
                \label{fig:ew_yields}
        \end{subfigure}~
                \begin{subfigure}[b]{0.4\textwidth}
                \includegraphics[width=\textwidth]{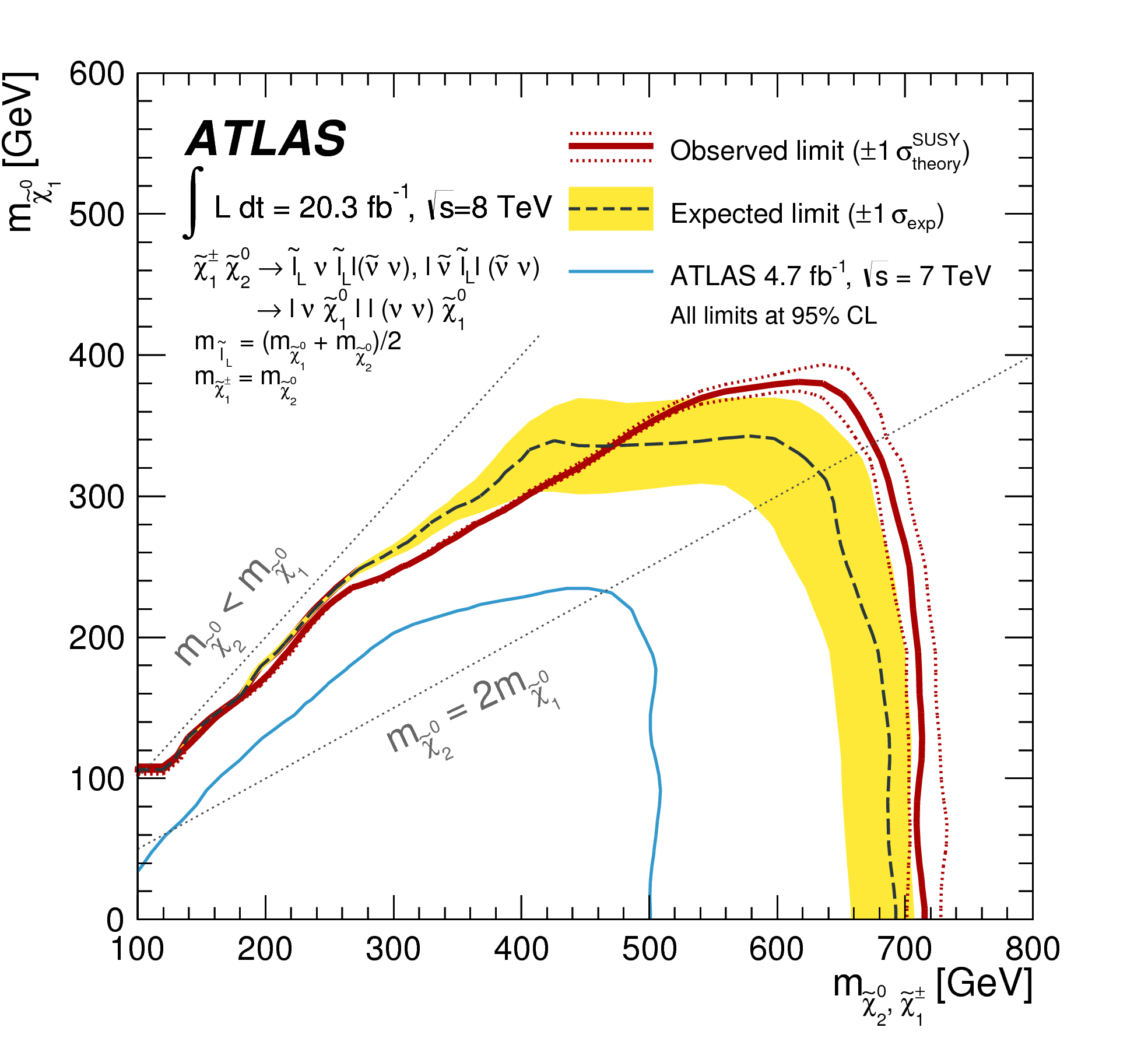}
                \caption{}
                \label{fig:ew_slep}
        \end{subfigure}
     \\%
        \begin{subfigure}[b]{0.4\textwidth}
                \includegraphics[width=\textwidth]{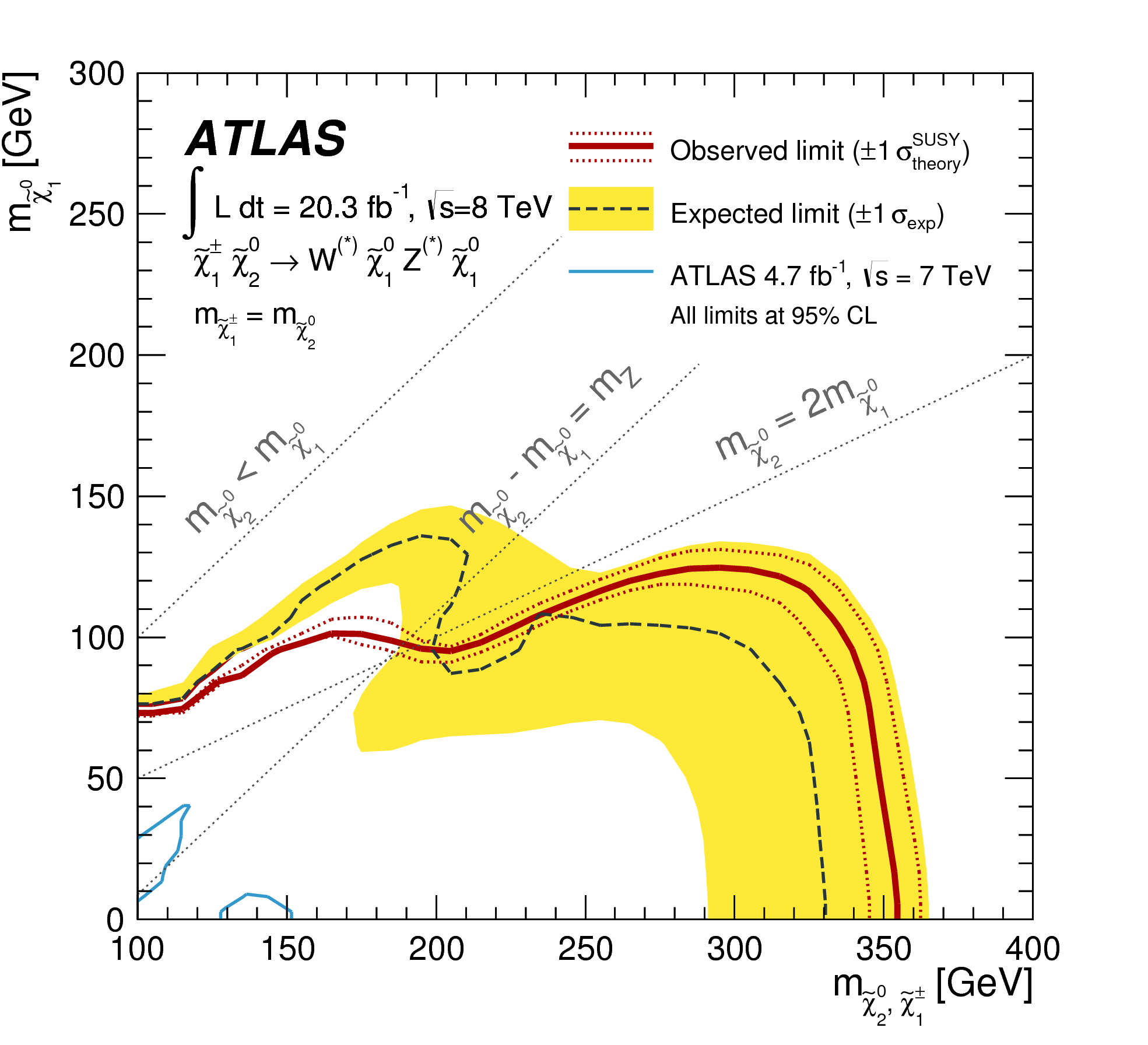}
                \caption{}
                \label{fig:ew_bos}
        \end{subfigure}        ~ 
        \begin{subfigure}[b]{0.4\textwidth}
                \includegraphics[width=\textwidth]{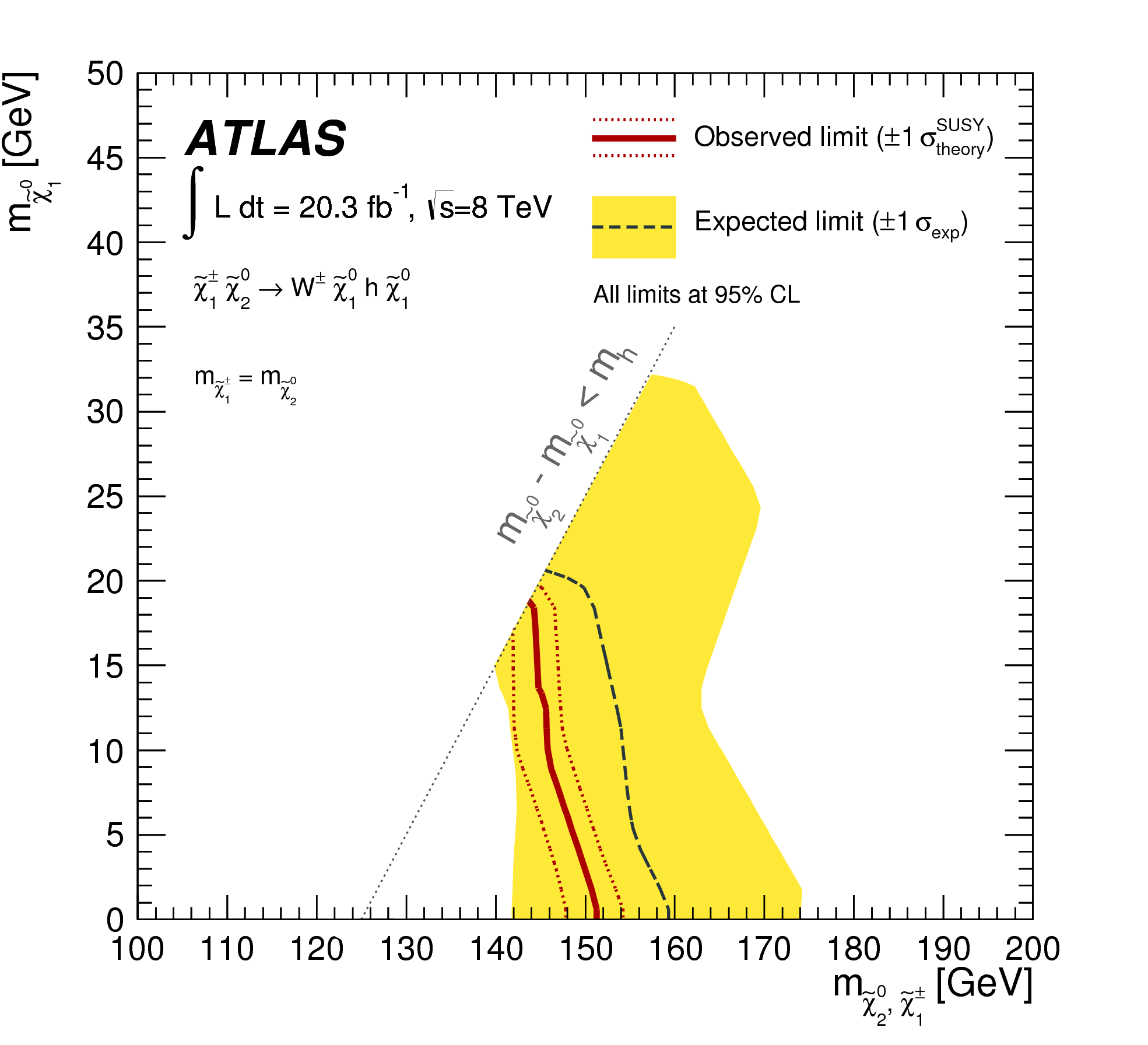}
                \caption{}
                \label{fig:ew_higgs}
        \end{subfigure}
        \caption{(a) Expected distributions of SM background events and observed data distributions in the 3 lepton (electrons and muons) signal region. (b)-(d) Limits set on the chargino - neutralino pair production in different decay scenarios: (b) via sleptons, (c) via WZ and (d) via Higgs. \cite{Aad:2014nua} }\label{fig:ew}
\end{figure}

\begin{figure}
\begin{center}
\includegraphics[width=0.6\columnwidth]{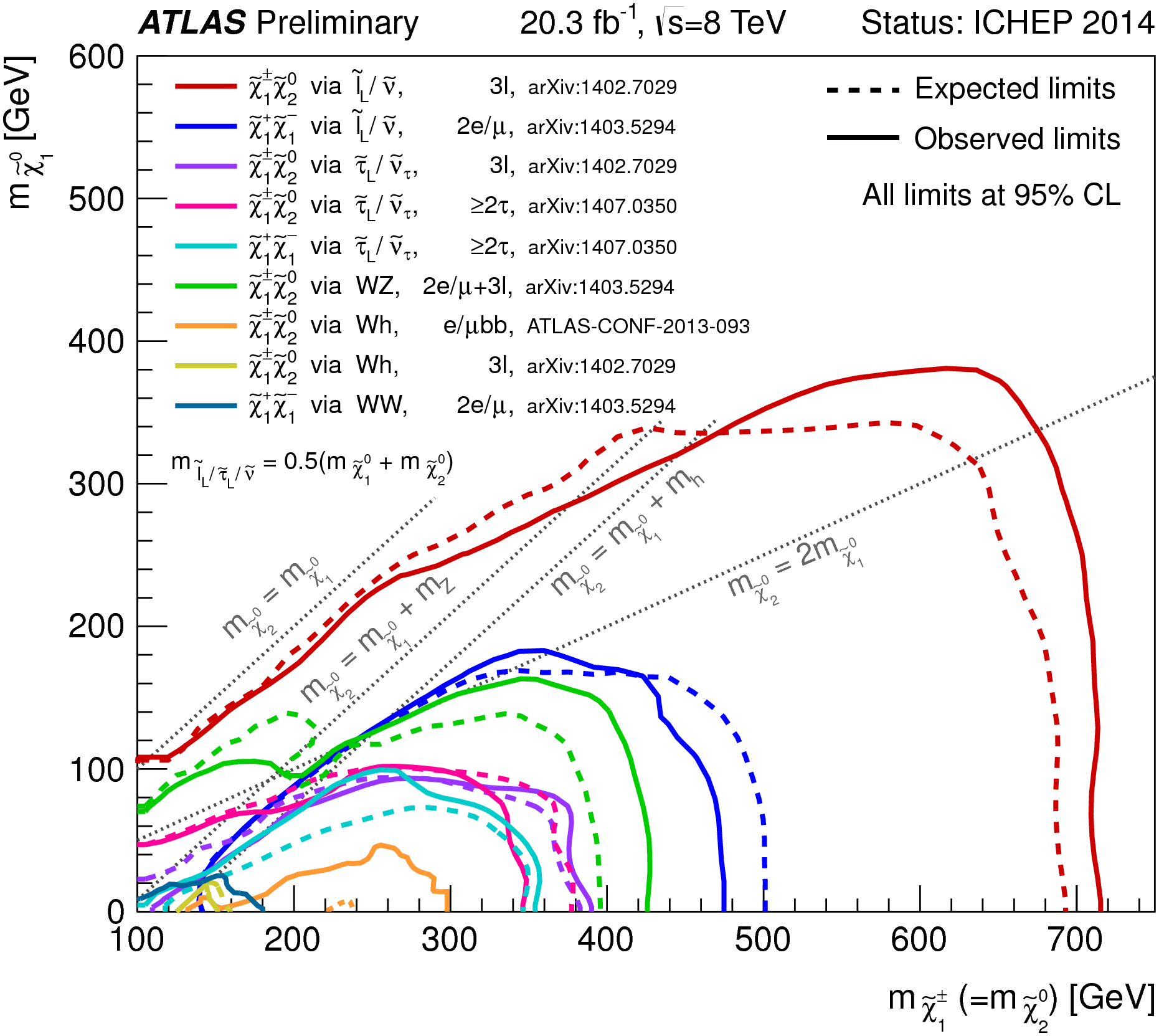}
\caption{Summary of ATLAS search for electroweak production of charginos and neutralinos. Limits are set in the chargino neutralino mass planes. The dashed and solid lines show the expected and observed limits, respectively, including all uncertainties except the theoretical signal cross section uncertainties. \cite{ATLAS:summaryplots} }
\label{fig:example3}
\end{center}
\end{figure}

The  $\tilde\chi_1^\pm$ and $\tilde\chi_2^0$, assumed to be produced in pairs,   decay to the LSP as  follows: $\tilde\chi_1^\pm \rightarrow W^\pm(\rightarrow l\nu) \tilde\chi_1^0$ and $\tilde\chi_2^0 \rightarrow Z(\rightarrow ll) \tilde\chi_1^0$. In the case of the $\tilde\chi_2^0\tilde\chi_1^\pm$ production the signal sensitivity is illustrated in Fig. \ref{fig:ew_yields} where the 3 lepton phase space is split into 20 bins to increase the sensitivity to different possible models\cite{Aad:2014nua}.  The bins are defined using a set of requirements on the invariant mass of opposite charged leptons and transverse mass ($m_{T2}$) of the third lepton.  Chargino masses above 350 GeV are excluded for the decay via SM bosons and 700 GeV for the decays via sleptons (see Fig.~\ref{fig:ew_slep}). For the decay $\tilde\chi_2^0 \rightarrow h \tilde\chi_1^0$ a weaker limit on the chargino mass of 150 GeV is set \cite{Aad:2014nua}. The summary of all electroweakinos ATLAS searches can be seen in Fig. \ref{fig:example3}.

\subsection{R-parity violating scenarios}
\label{sec:longlived}

\begin{figure}
        \centering
        \begin{subfigure}[b]{0.45\textwidth}
                \includegraphics[width=\textwidth]{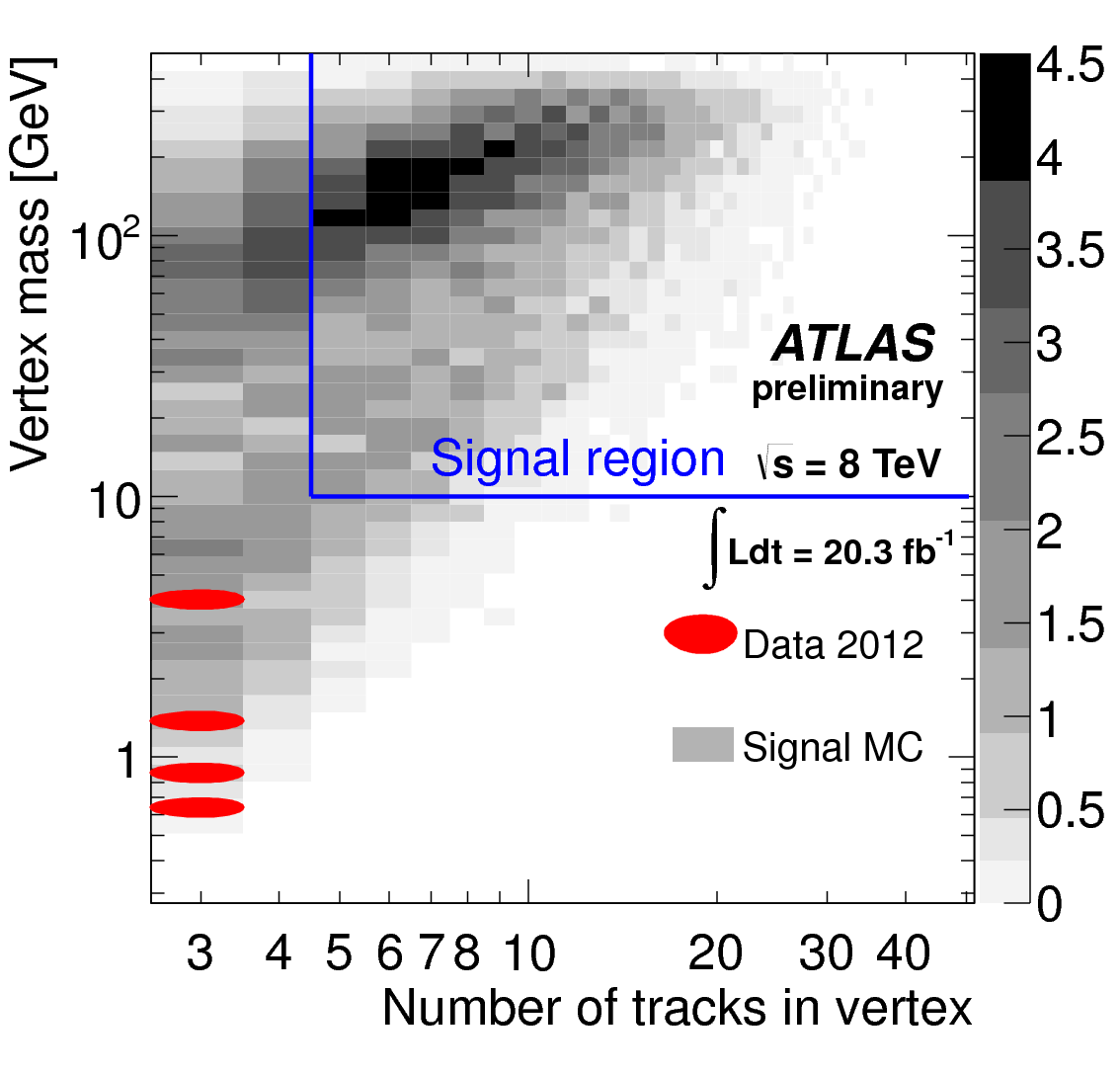}
                \caption{}
                \label{fig:longlived_sig}
        \end{subfigure}
                \begin{subfigure}[b]{0.45\textwidth}
                \includegraphics[width=\textwidth]{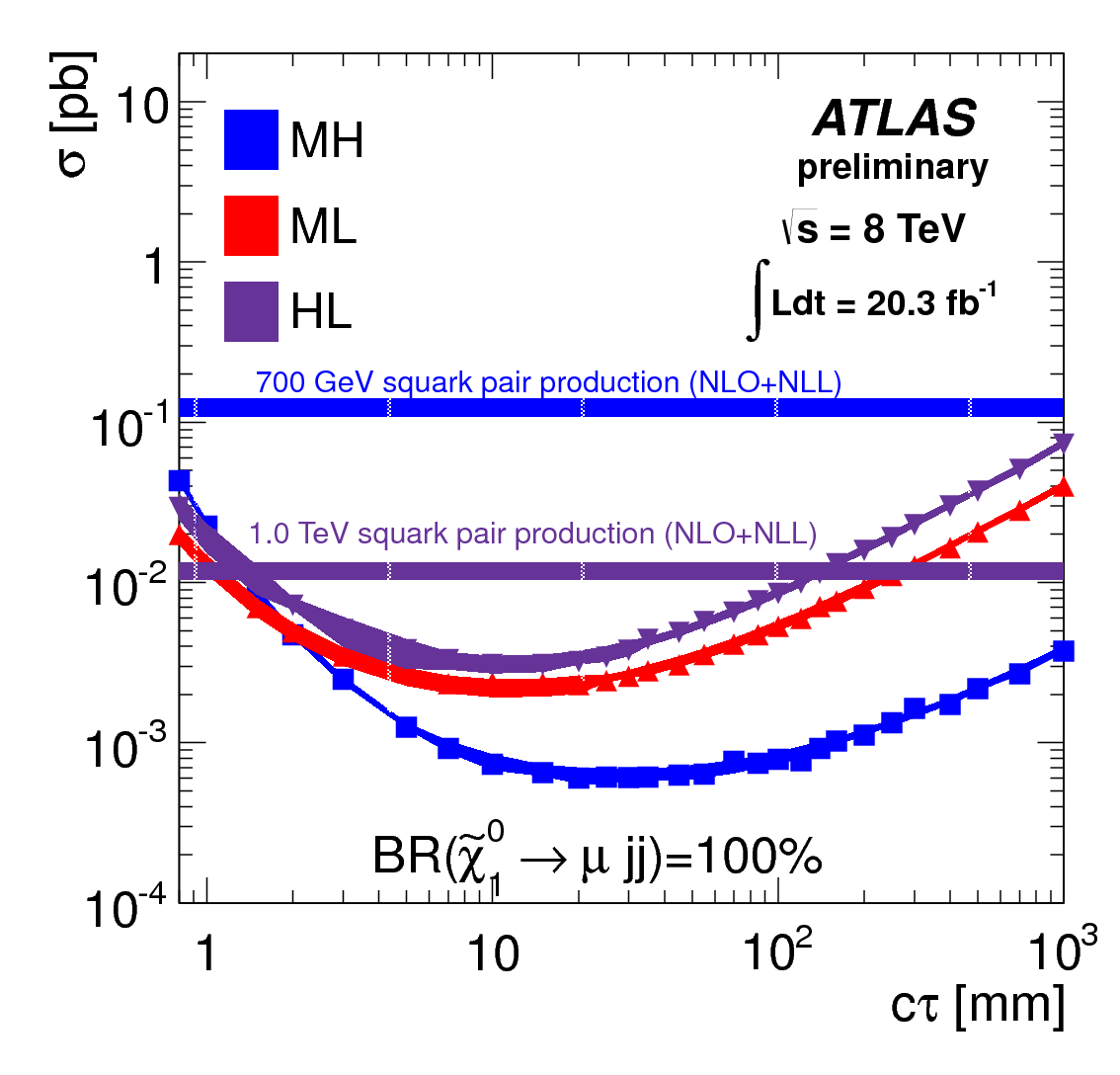}
                \caption{}
                \label{fig:longlived_xsec}
        \end{subfigure}%
                \caption{(a) Vertex mass versul vertex track multiplicity. The region where the signal is on the upper right side with no events observed in data.   (b) Upper limits at 96\% CL on the cross-section versus neutralino lifetime for different combinations of squark and neutralino masses.  \cite{ATLAS:Conf-2013-092}}\label{fig:squarks_gluino}
\end{figure}

In the case of R-parity violation the signatures can be spectacular. For example one can have an RPV-violating coupling that allows for  the decay of the neutralino to a muon and two quarks. This can produce a specific signature of a displaced vertex with high track multiplicity, among the tracks being one associated with a muon \cite{ATLAS:Conf-2013-092}. For such a model you can see  in Fig. \ref{fig:longlived_sig} the vertex mass versus the vertex tracks multiplicity for the expected signal and data. Upper  limits are set on different scenarios of neutralino and squark mass and different lifetimes of the neutralino and can be seen in Fig. \ref{fig:longlived_xsec}. 

\section{Summary}

The ATLAS collaboration has performed a wide and thorough search for a multitude of possible SUSY scenarios. Despite all the efforts no sign of SUSY was observed. Limits on the masses of the SUSY particles were set in different scenarios and being of about 1 TeV for squarks and gluinos and few hundreds of GeV for the other particles. The full list of results of searches for SUSY in ATLAS can be found in \cite{ATLAS:susyresults} . With the higher energy of the LHC in the run starting in 2015 a new phase space for SUSY searches is opened. The experience of searches presented here  will provide the stepping stone for the upcoming ones. With the new data  evidence for SUSY will be discovered or the current limits will be pushed to higher masses.   

\end{document}